\documentclass[prb,twocolumn,amsmath,amssymb]{revtex4}% Physical Review B
\usepackage{graphicx}
\usepackage{dcolumn}% Align table columns on decimal point
\usepackage{bm}% bold math
\usepackage{amssymb}

\begin{document}

\title{High charge-carrier mobility and low trap density in a rubrene derivative}

\author{S.~Haas}
\author{A.~F.~Stassen}
\author{G.~Schuck}
\author{K.~P.~Pernstich}
\author{D.~J.~Gundlach}
\author{B.~Batlogg}
\email{batlogg@phys.ethz.ch}
\affiliation{Laboratory for Solid State Physics, ETH Zurich, 
8093 Zurich, Switzerland
}
\author{U.~Berens}
\author{H.-J.~Kirner}
\affiliation{Ciba Specialty Chemistry Inc., 4002 Basel, Switzerland }

\date{\today}
%\maketitle

\begin{abstract}
We have synthesized, crystallized and studied the structural and electric transport properties of organic molecular crystals based on a rubrene derivative with {\em t\/}-butyl sidegroups at the 5,11 positions.
Two crystalline modifications are observed: one (A) distinct from that of rubrene with larger spacings between the naphtacene backbones, the other (B) with a in-plane structure presumably very similar compared to rubrene. 
The electric transport properties reflect the different structures:
in the latter phase (B) the in-plane hole mobility of 12\,cm$^2$/Vs measured on single crystal FETs is just as high as in rubrene crystals, while in the A phase no field-effect could be measured.
The high crystal quality, studied in detail for B, reflects itself in the density of gap states: The deep-level trap density as low as $10^{15}$ cm$^{-3}$ eV$^{-1}$ has been measured, and an exponential band tail with a characteristic energy of 22 meV is observed. The bulk mobility perpendicular to the molecular planes is estimated to be of order of $10^{-3}$ -- $10^{-1}$ cm$^2$/Vs.
\end{abstract}

\maketitle

\section{Introduction}

Unsubstituted, linear acenes have been at the center of organic molecular crystal research for the past decades due to their model character for the study of electric and optical properties \cite{Karl2003}, and due to the promise of pentacene thin films as a high-performance organic thin film transistor material \cite{Kelley2003}. Recently, rubrene has joined this selection as several groups have reported a charge mobility in field effect devices near or above 10 cm$^2$/Vs \cite{Podzorov2004,de-Boer2004,claudia04,Sundar2004,Stassen2004}.
Unlike pentacene, which is one of the small molecules of choice for thin film applications, rubrene does not readily form crystalline ordered films, neither by evaporation nor by solution-based deposition. Recently, however, ordered rubrene films were produced using alternative methods: evaporation onto a pentacene film as substrate \cite{Haemori2005}, and by incorporation of rubrene into a polymer matrix \cite{Stutzmann05}.
From a technological standpoint, it is desirable to synthesize derivatives of rubrene that would easily form high-quality thin films, with the promising electrical properties of unmodified rubrene.

It is commonly assumed, and supported by calculations, that an increased intermolecular $\pi$-orbital overlap increases the bandwidth, and thus the mobility \cite{pope}.
Therefore structures with $\pi$-stacking rather than the prevalent herringbone packing are expected to show higher mobilities \cite{Curtis2004}. Accordingly, several new materials have been chosen or designed. Successful  examples are the functionalized pentacene derivatives \cite{Anthony2001}, which show as solution-deposited films field-effect mobilities as high as 1
cm$^2$/Vs \cite{Payne2005a}. Also rubrene exhibits a slip-stack packing with efficient $\pi$-overlap.
Recently, a direct correlation between structure and mobility has been reported for various tetrathiafulvalene derivatives, crystallizing in three types of structures \cite{Mas-Torrent2004}. In contrast, almost no effect of (small) changes in the packing on the mobility has been observed for various tetracene derivatives \cite{TcD}. By adding side-groups to rubrene, leaving the $\pi$-system of the naphthacene backbone nearly unchanged but altering the molecular packing, our studies aim at a structure--mobility relationship for rubrene and derivatives. 

In the present work, we have synthesized a modification of rubrene, grown single crystals, studied the crystal structure, and measured the electric properties. Of the two polymorphs grown, one (A) has a structure drastically different from the packing of unsubstituted rubrene, with a strongly twisted naphthacene backbone and enhanced spacing between them. No field-effect is observed with crystals of polymorph A. On the other hand, polymorph B shows a field-effect mobility as high as that of rubrene, indicating a similar in-plane packing. Because the crystals of polymorph B grow as very thin platelets only, the full structure is not determined, but the $00l$ reflections indicate a molecule packing analogous to an other high-mobility rubrene derivative.
The density of electronic states (DOS) in the band gap has been measured using the method of temperature-dependent space-charge limited current (TD-SCLC) spectroscopy. The crystals (B) are of very high electric quality with trap densities as low as $\sim$$2\times10^{15}$ cm$^{-3}$eV$^{-1}$ at $\sim$0.2\,eV from the mobility edge, and a steep exponential rise associated with band tail states on approaching the band edge.

\section{Experimental Section}\label{exp}

\begin{figure}[t!]
\includegraphics[width=0.9\columnwidth]{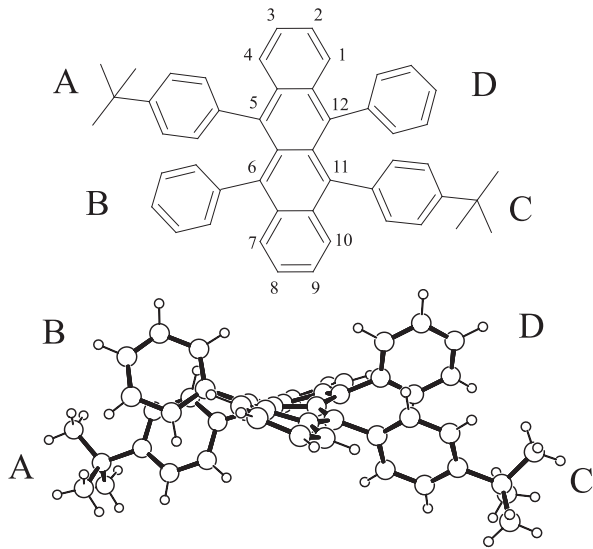}
\caption{\label{strichformel}
Rubrene with side-groups: 5,11-bis-(4-{\it tert}-butyl-phenyl)-6,12-diphenyl-naphthacene, 
C$_{50}$H$_{44}$. The view along the long axis of the naphthacene backbone reveals the large twist of 43 degrees in its polymorphic form A (non conducting). Structure drawings made with Ortep-3 for Windows \cite{Farrugia1997}.}
\end{figure}

Single crystals of 5,11-bis-(4-{\it tert}-butyl-phenyl)-6,12-diphenyl-naphthacene [bis-(5,11-para-{\it t}-butyl)rubrene, {\bf 5,11-BTBR}, C$_{50}$H$_{44}$, see
Figure~\ref{strichformel}a), synthesis according to Ref.~\onlinecite{Kopranenkov1972}] have been grown by physical vapor transport \cite{Kloc1997,Laudise1998} at 260\,$^{\circ}$C, using high purity argon as the 
transport gas. The crystals are transparent orange colored platelets, typically 0.1--2 $\mu$m thick.
Only at slightly higher temperature (and with longer growth time), a few bulky crystals could be grown for full structural characterization. As discussed below, these crystals are of one of the two polymorphs, and their structure is analized in detail by XRD.

Field-effect transistor (FET) measurements and TD-SCLC spectroscopy were used to gain insight to the electronic properties of this material.
SCLC was measured perpendicular to the surface of the 5,11-BTBR platelets in a
sandwich-type sample layout (typical cross section $\sim 1.5\cdot 10^{-5}$\,cm$^2$) with bottom electrodes (Au/Cr) evaporated on a glass substrate, and a Au top electrode evaporated directly onto the crystal. 
FETs were fabricated in a flip crystal technique \cite{Takeya2003}, where source and drain electrodes (Au) were deposited on the surface of an oxidized Si wafer, and the crystal is carefully placed on these contacts and sticks to the substrate due to electrostatic adhesion. Beforehand, a monolayer of OTS was applied to the substrate to improve the device performance \cite{Gundlach2001a}. Typical device dimensions were L=100\,$\mu$m and W=200--800\,$\mu$m.
More details about the measurement techniques are described elsewhere for FETs \cite{claudia04} and SCLC \cite{Krellner}.
To exclude environmental influence, all electrical measurements were performed in a 
helium atmosphere.

\section{Results and Discussion}\label{results}

\begin{figure*}[tbh]
\includegraphics[angle=-90,width=1.7\columnwidth]{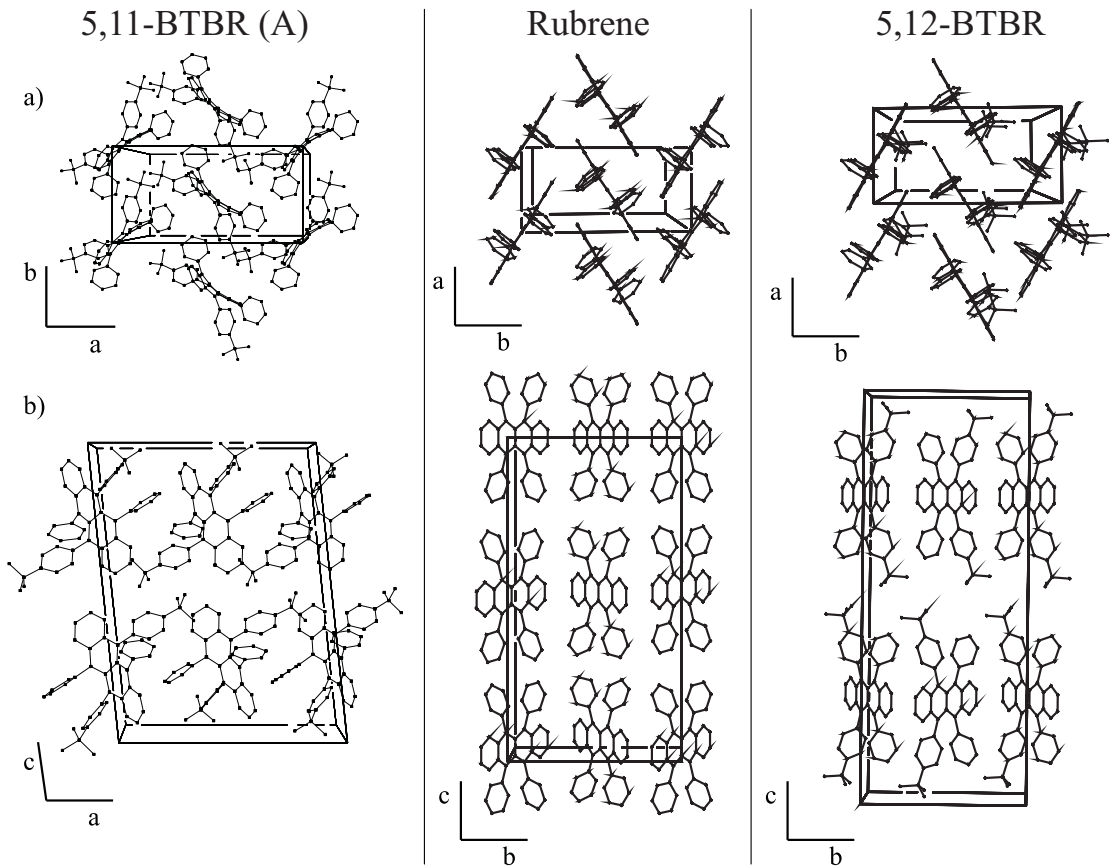}
\caption{\label{Strukturen}Perspective views of the structures of 5,11-BTBR (polymorph A, left), rubrene \cite{Bulgarovskaya83} (center), and 5,12-BTBR (similar to 5,11-BTBR B, right). For clarity, only one layer of molecules with respect to the paper plane is shown. 
a) View onto the $a$-$b$ face: 
In rubrene and 5,12-BTBR (5,11-BTBR B), $\pi$-stacking between the molecules is established along $a$ (short axis), whereas in 5,11-BTBR (A) the spacing between the backbones is twice as wide in the equivalent direction of $b$.
b) View along $a$ or $b$, respectively, illustrates the molecular layers. 
In the single crystal FETs the current flows along the layers in the $a$-$b$ direction (in-plane), while in the TD-SCLC measurements the current flows in direction of $c$ or $c^{\star}$, respectively, i.e. perpendicular to $a$,$b$.}
\end{figure*}

\subsection{Structure}
Interestingly, two polymorphs, A and B, have been identified, which differ in their $d$-spacing perpendicular to the crystal platelets: 23.4\,\AA\ and 35.1\,\AA. In the following, the structure information of the two polymorphs are compared.

Details of the structure of polymorph A are published elsewhere\cite{schuck}.
Remarkably, the naphthacene backbone of the molecules is significantly twisted in polymorph A, with a twist angle of 43 degrees between the two opposite C-C bonds at both ends of the backbone (c.f. Fig.~\ref{strichformel}b).
Figure \ref{Strukturen} shows the molecular packing of 5,11-BTBR (A). The molecules form a layered structure, similar to linear acenes such as pentacene \cite{Mattheus2001} with the naphthacene backbone standing upright. 
It has a higher symmetry ($P2_1/a$) compared to e.g. tetracene and pentacene ($P\bar{2}$), and four molecules are in the unit cell (Z=4).
The naphthacene backbones are further apart with in-plane lattice constants of $a$=17.76\,\AA\  and $b$=9.024\,\AA, compared with the in-plane axes of roughly 8\,\AA $\times$6\,\AA\ for 
the unsubstituted acenes. With these large in-plane spacings, the in-plane arrangement differs from the classical herringbone structure, resembling slip-stack structure type, albeit without short-distance interactions enabling $\pi$-stacking (see Fig. \ref{Strukturen}a).

In contrast, the (unmodified) rubrene molecules have a nearly perfectly planar naphthacene backbone, and the arrangement of the molecules differs from the classical herringbone structure: The {\em long} axis of the naphthacene backbone lies {\em in\/} the molecular planes, enabling $\pi$-stacking in direction of the $a$-axis \cite{RuD_RubreneAchsen}, as depicted in Fig.~\ref{Strukturen}. (Crystallographic data for rubrene: see Ref. \onlinecite{Bulgarovskaya83}.)

The material seems to exclusively grow as ultra-thin platelets, therefore the full structure of polymorph B could not be solved so far. From measurements of the $d$-spacing perpendicular to the extended crystal surface, we can assume a structure closely related to the one found for a constitutional isomer, 5,12-BTBR \cite{schuck1, Arno_RuD, RuD_NMR}. In the case of 5,12-BTBR, the in-plane arrangement of the molecules is very similar to that of rubrene, with even shorter distances between the naphthacene backbones (3.55\,\AA\ compared to  3.74\,\AA). However, the addition of the {\em t-\/}butyl groups increases the inter-layer spacing by 31\,\%. Interestingly, it leaves the backbone almost perfectly planar (see Fig.~\ref{Strukturen}).

In FET measurements the current flows within the $a$,$b$-plane, while in SCLC it flows 
perpendicular to the molecular layers of 5,11-BTBR and rubrene, as indicated in Fig.~\ref{Strukturen}.

\subsection{Field-effect transistor measurements}\label{FETchapter}

\begin{figure}[htb]
\includegraphics[width=0.9\columnwidth]{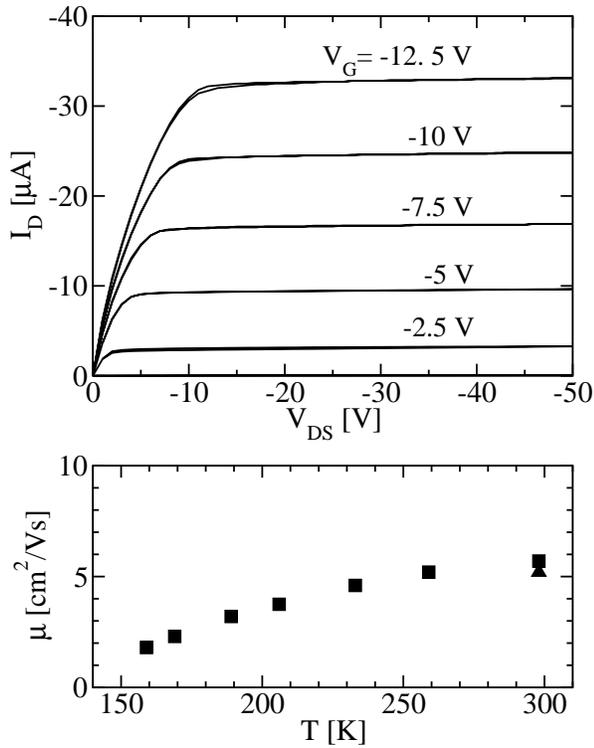}
\caption{\label{FET} Single-crystal field-effect-transistors:
 The upper panel shows the output characteristic of a flip-crystal FET measured
at room temperature. (Dimensions: $W$=770$\mu$m, $L$=100\,$\mu$m, $d_{\rm SiO_2}$=300\,nm.)
Lower panel: Mobility $\mu$ for a flip crystal FET as a function of temperature, indicating a slight decrease from 6 cm$^2$/Vs at room temperature to 2 cm$^2$/Vs at 160\,K.}
\end{figure}

A typical output characteristic for a 5,11-BTBR (B) FET is shown in Fig.~\ref{FET} (upper panel.) The observed hysteresis is very small, indicating that trapping/releasing effects between two successive measurement sweeps over 20--30 seconds are negligible. The turn-on voltage $V_{\rm on}$ and threshold $V_t $ are below 8\,V and 5\,V, respectively, in typical samples. The mobilities were calculated using the standard MOSFET equation for the drain current in the saturation regime at an effective gate voltage of -10\,V. In all devices with OTS-treated oxide, the mobility exceeds 1\,cm$^2$/Vs, and a maximum mobility of 12\,cm$^2$/Vs was observed in the best sample.  
Temperature dependent measurements have been performed on a representative device. The resulting 
values for the mobility are shown in Fig.~\ref{FET} (lower panel). We note a small decrease of $\mu$ (from 5.7\,cm$^2$/Vs to $\sim$1.8\,cm$^2$/Vs) when cooling from room temperature to 160\,K. $V_{\rm on}$ and $V_{\rm t}$ also decrease from 8\,V to 5\,V and 4\,V to 1.2\,V, respectively. After warming-up to room temperature, the measured characteristics as well as the mobility are within 10\,\% of the original measurements. No change of the performance is observed after storing the device in inert atmosphere for one month.

In rubrene the distance between the positions of the C-atoms of two backbones is as short as 3.75\,\AA\ (edge molecule to middle one), and 3.85\,\AA\ (edge to edge molecule, zigzag along $a$). Taking the known structure of 5,12-BTBR as a model for the structure of 5,11-BTBR (B), we can expect in-plane backbone-backbone distances comparable to rubrene, i.e. 3.6\,\AA\ to 3.9\,\AA.
As the relevant bandwidth in a crystal is determined by the details of the HOMO and LUMO wave functions, a detailed electronic structure calculation is needed to quantify small differences in band structure associated with variations in packing geometry (e.g. pentacene \cite{Tiago2003, Hummer2005}).

The shortest backbone-backbone distance in 5,11-BTBR (A) is not shorter than 6.5\,\AA, which is commonly expected to drastically reduce the $\pi$-$\pi^{\star}$ overlap and thus the bandwidth. Consequently, we are not able to measure any field-effect mobility in 5,11-BTBR (A).
Norton and Houk \cite{Norton2005} have calculated that the twist of acenes (Anthracene--Heptacene) basically doesn't change the HOMO-LUMO energies, i.e. the energy gap is not affected. On the other hand, too large a twist will affect the aromaticity of the molecules, increasing the localization of charge on the molecule. Experimentally, a twist of 144$^{\circ}$ has been observed in a pentacene derivative \cite{Lu2004}.

\subsection{Trap density-of-states measurements by TD-SCLC}\label{TDSCLC}

\begin{figure}[htb]
\includegraphics[width=\columnwidth]{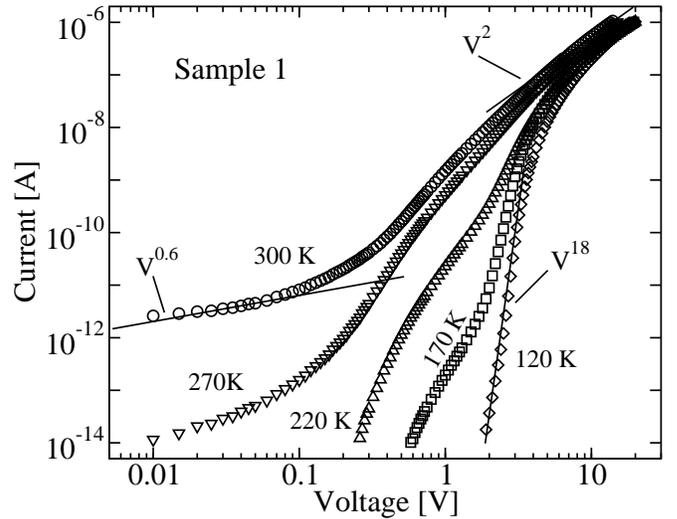}
\caption{\label{SCLC} Space charge limited current at various temperatures for sample 1. At room temperature, three regions can be discerned: A (sub-)ohmic behavior, dominated by contact effects (for $V\le0.3$\,V), then a steep increase, and finally a turn-over to a constant slope of 2 in the $\log$-$\log$ plot.
For lower temperatures, $V^2$ behavior is still reached, after a very steep increase proportional max. $V^{18}$. The $V^2$ behavior is due to the quasi-Fermi level moving in the exponential part of the density of states (``band tail states''), but the trap-free limit is not yet reached.
For the extraction of the DOS($E$\/), $I$-$V$ curves measured between 130\,K and 180\,K have been used ($\Delta T=10$\,K.)}
\end{figure}

In the previous section, the performance of 5,11-BTBR (B) devices in terms of high field-effect mobilities has been presented. Another important measure for the intrinsic electronic quality is the density of in-gap (trap) states, since electrical transport is known to be severely affected by charge carrier trapping even in the best organic crystals. To quantify the DOS, we apply the method of temperature-dependent space-charge limited current (TD-SCLC) spectroscopy \cite{Schauer04,Krellner} to crystals of 5,11-BTBR (B).

The DOS is reflected in the shape of the measured $I$-$V$ curves.
{\em Temperature-dependent\/} measurements are needed: 1) as an appropriate method to assess the possible influence of contacts at low voltages,  and 2) in order to associate a given applied voltage with the corresponding distance of the quasi-Fermi level from the band mobility edge, which is done by measuring the activation energy $E_{\rm A}$(V).
As the Fermi level is moved toward the band edge with increasing voltage, $E_{\rm A}$ has to decrease monotonically with increasing voltage. Contact-limited current is thus recognized as a deviation from this monotonic dependence of $E_{\rm A}$ upon $V$. Additionally, abrupt changes in the density of states are recognized, since e.g. discrete trap levels lead to a pinning of $E_{\rm A}$. Because of the asymmetry of DOS($E$) around $E_{\rm F}$, $E_{\rm A}$ is corrected to $E_{\rm D}$ (dominant energy), defined by the statistical shift.

The current-voltage characteristics measured at room temperature for several samples of 5,11-BTBR (B) have in common, that a pronounced, steep increase of the current occurs at relatively low voltage, indicating the trap-filling SCLC region. Additionally, a gradual transition to $I\propto V^2$ is observed for most samples. This is a first evidence for a low over-all trap density. The observation of $I\propto V^2$ dependence, however, does not, by itself, indicate that the trap free range has been reached. A quantitative analysis of the DOS with energy resolution $\sim$$kT$, for instance by the means of TD-SCLC, is therefore still needed.
Due to thermo-mechanical stress during temperature cycles in the course of the TD-SCLC measurements, several crystals suffered from cracks, interrupting the top electrode or directly affecting the measurement cross-sections. Two complete sets of data are discussed here (sample 1: $d$=1.25\,$\mu$m, sample 2: $d$=0.4\,$\mu$m, $A$=$1.5\cdot 10^{-5}$\,cm$^2$ for both samples.)

The current-voltage characteristics for sample 1 for selected temperatures are plotted in Figure~\ref{SCLC}. The curve measured at 300K shows three distinct regions: (sub)ohmic current at low voltage, trap filling SCLC, and apparently ``trap-free'' SCLC at highest voltage. For low temperatures, only the trap filling and the ``trap-free'' region are observed, because the ohmic current is too small to be measured with our measurement setup. 
 At low temperatures, the increase in current during trap filling is very steep with slopes of up to 18 in the $\log(I)$-$\log(V)$-plot.

\begin{figure}[t!]
\includegraphics[width=0.9\columnwidth]{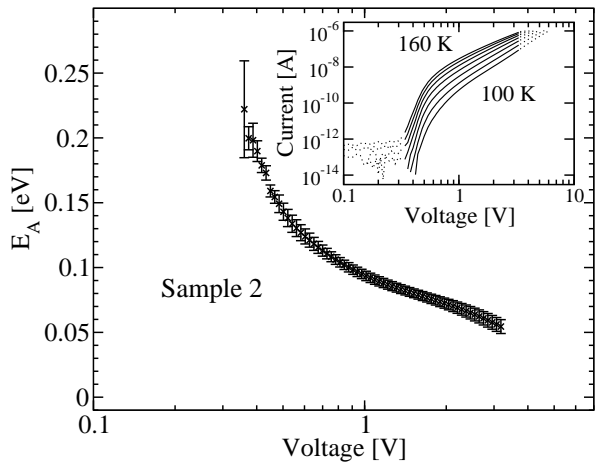}
\caption{\label{EASample2} Activation energy $E_{\rm A}(V)$ for sample 2. 
Inset: $I$-$V$ curves for sample 2. Below 0.34 V, where the current is contact-limited, as well as above 3.3 V, close to the rollover to a $I$$\propto$$V^2$ behavior, the current is not thermally activated. Thus meaningful activation energies can be extracted only in this intermediate range. Consequently, only data represented by solid lines was used for the extraction of $E_{\rm A}(V)$ and the DOS($E$\/).}
\end{figure}

Arrhenius-plots of the temperature dependent data reveal thermally activated
behavior for $\sim$0.2\,V$\le$$U$\/$\le$5\,V (sample 1), and for 0.34\,V$\le$$U$\/$\le$3.3\,V (sample 2). At lower voltages, both Arrhenius plots and subohmic $I$-$V$ characteristics indicate a current limitation by the contact. At highest voltage, $\log(I)$ does not depend linearly on $1/T$, as the quasi Fermi level moves within $\sim$$kT$ to the mobility edge.
Thus extraction of $E_{\rm A}$ is limited to this intermediate voltage range (cf. Fig.~\ref{EASample2} for sample 2).

\begin{figure}[t!]
\includegraphics[width=0.9\columnwidth]{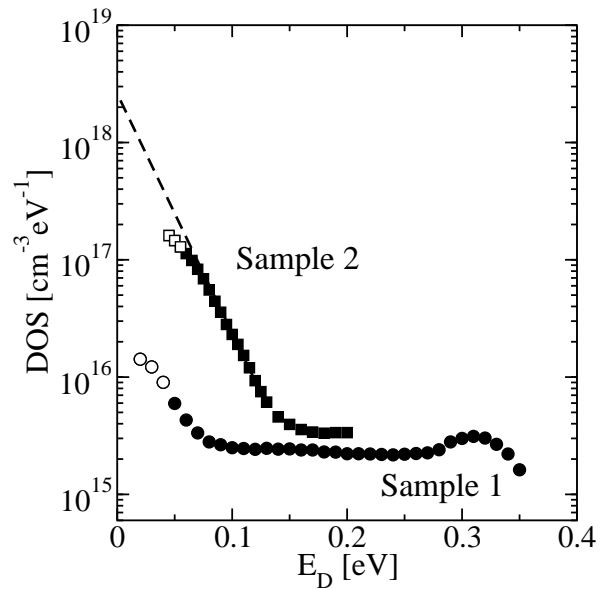}
\caption{\label{DOS} The density of states toward midgap is in the range of $10^{15}$\,cm$^{-3}$eV$^{-1}$. Close to the band edge, the DOS increases in a way reminiscent of band tail states (sample 2). Within 2--3kT to the mobility edge, the analysis starts to fail, causing an unphysical roll-over of the DOS (open symbols).}
\end{figure}

The extracted DOS for both samples is shown in Figure~\ref{DOS}. In the range from 0.15 to 0.3\,eV, the over-all density of trap states is rather low, in the range of $\sim$$10^{15}$ cm$^{-3}$eV$^{-1}$. On approaching the mobility edge ($E$=\,0), tail-like states with a characteristic energy of 22 meV (sample 2), and possibly the onset of a band tail for sample 1 emerge. Too close to the band, where $E_{\rm D}$ becomes comparable to $kT$, the analysis procedure fails, leading to an unphysical roll-over of the DOS (open symbols in Fig.~\ref{DOS}).
The DOS is as low as in the best rubrene samples, and the widths of the band tails are very similar to the ones measured in rubrene \cite{Krellner}. Worth mentioning is the fact that the DOS reported here is several orders of magnitude lower than the one for pentacene \cite{Lang02}, which was measured in a coplanar contact geometry which may emphasize the higher trap densities near/at the surface of the crystal.

A raw measure for the band mobility $\mu$ in 5,11-BTBR (B) perpendicular to the molecular layers (along the $c$-axis) is taken from extracted {\em effective\/} mobilities $\mu_{\rm eff}$, a purely numerical construct defined as the mobility at the band edge multiplied by the ratio of mobile to total injected charge. Extrapolated to the band edge ($E_{\rm D}$=\,0), $\mu_{\rm eff}\approx 10^{-3}$--$10^{-1}$\,cm$^2$/Vs in 5,11-BTBR (B).
A comparison with rubrene shows values in the same range for $\mu_{\rm eff}(E_{\rm D}$=\,0) perpendicular to the molecular layers \cite{diplom}, despite the enhanced inter-layer spacing in the derivative.

\section{Conclusions}

Single crystals of a rubrene derivative have been grown and the semiconducting polymorph B has been electrically characterized. The trap density in the energy gap is of the order of $10^{15}$\,cm$^{-3}$eV$^{-1}$; low enough to reveal band-tail like states with a characteristic energy of 22\,meV. This low DOS and the occurrence of {\em narrow\/} band tails are characteristic for high quality organic crystals. Presumably having a similar crystal structure and slip-stack in-plane arrangement of the molecules compared to rubrene, 5,11-BTBR (B) has shown a comparably high in-plane field-effect mobility of up to 12\,cm$^2$/Vs. 
It fits into the present understanding of the relation between structure and mobility, too, that the in-plane mobility is too low to be measured in 5,11-BTBR (A) because the naphthacene backbones are twice as far apart from each other as in rubrene, and $\pi$-stacking of the backbones is absent.

Several other modifications of rubrene are the subject of the ongoing research.
In order to get a more complete understanding of the relationship between crystal structure and charge transport, the full structure of 5,11-BTBR (B) and additional derivates, and, based thereon, band structure calculations are needed, since even small structural changes are expected to result in notable differences in wave function overlap. 

\section*{Acknowledgments}

We thank Oliver Dosenbach for the skillful assistance in the synthesis of the rubrenes, Claudia Goldmann and Wolfgang Kalb for fruitful discussions, and Kurt Mattenberger and Hanspeter Staub for technical support.


\begin{thebibliography}{36}
\expandafter\ifx\csname natexlab\endcsname\relax\def\natexlab#1{#1}\fi
\expandafter\ifx\csname bibnamefont\endcsname\relax
  \def\bibnamefont#1{#1}\fi
\expandafter\ifx\csname bibfnamefont\endcsname\relax
  \def\bibfnamefont#1{#1}\fi
\expandafter\ifx\csname citenamefont\endcsname\relax
  \def\citenamefont#1{#1}\fi
\expandafter\ifx\csname url\endcsname\relax
  \def\url#1{\texttt{#1}}\fi
\expandafter\ifx\csname urlprefix\endcsname\relax\def\urlprefix{URL }\fi
\providecommand{\bibinfo}[2]{#2}
\providecommand{\eprint}[2][]{\url{#2}}

\bibitem[{\citenamefont{Karl}(2003)}]{Karl2003}
\bibinfo{author}{\bibfnamefont{N.}~\bibnamefont{Karl}},
  \bibinfo{journal}{Synthetic Metals} \textbf{\bibinfo{volume}{133-134}},
  \bibinfo{pages}{649} (\bibinfo{year}{2003}).

\bibitem[{\citenamefont{Kelley et~al.}(2003)\citenamefont{Kelley, Muyres,
  Baude, Smith, and Jones}}]{Kelley2003}
\bibinfo{author}{\bibfnamefont{T.~W.} \bibnamefont{Kelley}},
  \bibinfo{author}{\bibfnamefont{D.~V.} \bibnamefont{Muyres}},
  \bibinfo{author}{\bibfnamefont{P.~F.} \bibnamefont{Baude}},
  \bibinfo{author}{\bibfnamefont{T.~P.} \bibnamefont{Smith}}, \bibnamefont{and}
  \bibinfo{author}{\bibfnamefont{T.~D.} \bibnamefont{Jones}}, in
  \emph{\bibinfo{booktitle}{Mat. Res. Soc. Symp. Proc. Vol. 771}}
  (\bibinfo{year}{2003}), p. \bibinfo{pages}{L6.5.1}.

\bibitem[{\citenamefont{Podzorov et~al.}(2004)\citenamefont{Podzorov, Menard,
  Borissov, Kiryukhin, Rogers, and Gershenson}}]{Podzorov2004}
\bibinfo{author}{\bibfnamefont{V.}~\bibnamefont{Podzorov}},
  \bibinfo{author}{\bibfnamefont{E.}~\bibnamefont{Menard}},
  \bibinfo{author}{\bibfnamefont{A.}~\bibnamefont{Borissov}},
  \bibinfo{author}{\bibfnamefont{V.}~\bibnamefont{Kiryukhin}},
  \bibinfo{author}{\bibfnamefont{J.~A.} \bibnamefont{Rogers}},
  \bibnamefont{and} \bibinfo{author}{\bibfnamefont{M.~E.}
  \bibnamefont{Gershenson}}, \bibinfo{journal}{Phys. Rev. Lett.}
  \textbf{\bibinfo{volume}{93(8)}}, \bibinfo{pages}{086602/1}
  (\bibinfo{year}{2004}).

\bibitem[{\citenamefont{de~Boer et~al.}(2004)\citenamefont{de~Boer, Gershenson,
  Morpurgo, and Podzorov}}]{de-Boer2004}
\bibinfo{author}{\bibfnamefont{R.~W.~I.} \bibnamefont{de~Boer}},
  \bibinfo{author}{\bibfnamefont{M.~E.} \bibnamefont{Gershenson}},
  \bibinfo{author}{\bibfnamefont{A.~F.} \bibnamefont{Morpurgo}},
  \bibnamefont{and} \bibinfo{author}{\bibfnamefont{V.}~\bibnamefont{Podzorov}},
  \bibinfo{journal}{Physica Status Solidi A} \textbf{\bibinfo{volume}{201(6)}},
  \bibinfo{pages}{1302} (\bibinfo{year}{2004}).

\bibitem[{\citenamefont{Goldmann et~al.}(2004)\citenamefont{Goldmann, Haas,
  Krellner, Pernstich, Gundlach, and Batlogg}}]{claudia04}
\bibinfo{author}{\bibfnamefont{C.}~\bibnamefont{Goldmann}},
  \bibinfo{author}{\bibfnamefont{S.}~\bibnamefont{Haas}},
  \bibinfo{author}{\bibfnamefont{C.}~\bibnamefont{Krellner}},
  \bibinfo{author}{\bibfnamefont{K.~P.} \bibnamefont{Pernstich}},
  \bibinfo{author}{\bibfnamefont{D.~J.} \bibnamefont{Gundlach}},
  \bibnamefont{and} \bibinfo{author}{\bibfnamefont{B.}~\bibnamefont{Batlogg}},
  \bibinfo{journal}{J. Appl. Phys.} \textbf{\bibinfo{volume}{96(4)}},
  \bibinfo{pages}{2080} (\bibinfo{year}{2004}).

\bibitem[{\citenamefont{Sundar et~al.}(2004)\citenamefont{Sundar, Zaumseil,
  Podzorov, Menard, Willett, Someya, Gershenson, and Rogers}}]{Sundar2004}
\bibinfo{author}{\bibfnamefont{V.~C.} \bibnamefont{Sundar}},
  \bibinfo{author}{\bibfnamefont{J.}~\bibnamefont{Zaumseil}},
  \bibinfo{author}{\bibfnamefont{V.}~\bibnamefont{Podzorov}},
  \bibinfo{author}{\bibfnamefont{E.}~\bibnamefont{Menard}},
  \bibinfo{author}{\bibfnamefont{R.~L.} \bibnamefont{Willett}},
  \bibinfo{author}{\bibfnamefont{T.}~\bibnamefont{Someya}},
  \bibinfo{author}{\bibfnamefont{M.~E.} \bibnamefont{Gershenson}},
  \bibnamefont{and} \bibinfo{author}{\bibfnamefont{J.~A.}
  \bibnamefont{Rogers}}, \bibinfo{journal}{Science}
  \textbf{\bibinfo{volume}{303(5664)}}, \bibinfo{pages}{1644}
  (\bibinfo{year}{2004}).

\bibitem[{\citenamefont{Stassen et~al.}(2004)\citenamefont{Stassen, de~Boer,
  Iosad, and Morpurgo}}]{Stassen2004}
\bibinfo{author}{\bibfnamefont{A.~F.} \bibnamefont{Stassen}},
  \bibinfo{author}{\bibfnamefont{R.~W.~I.} \bibnamefont{de~Boer}},
  \bibinfo{author}{\bibfnamefont{N.~N.} \bibnamefont{Iosad}}, \bibnamefont{and}
  \bibinfo{author}{\bibfnamefont{A.~F.} \bibnamefont{Morpurgo}},
  \bibinfo{journal}{Appl. Phys. Lett.} \textbf{\bibinfo{volume}{85(17)}},
  \bibinfo{pages}{3899} (\bibinfo{year}{2004}).

\bibitem[{\citenamefont{Haemori et~al.}(2005)\citenamefont{Haemori, Yamaguchi,
  Yaginuma, Itaka, and Koinuma}}]{Haemori2005}
\bibinfo{author}{\bibfnamefont{M.}~\bibnamefont{Haemori}},
  \bibinfo{author}{\bibfnamefont{J.}~\bibnamefont{Yamaguchi}},
  \bibinfo{author}{\bibfnamefont{S.}~\bibnamefont{Yaginuma}},
  \bibinfo{author}{\bibfnamefont{K.}~\bibnamefont{Itaka}}, \bibnamefont{and}
  \bibinfo{author}{\bibfnamefont{H.}~\bibnamefont{Koinuma}},
  \bibinfo{journal}{Jpn. J. Appl. Phys.} \textbf{\bibinfo{volume}{44}},
  \bibinfo{pages}{3740} (\bibinfo{year}{2005}).

\bibitem[{\citenamefont{Stingelin-Stutzmann
  et~al.}(2005)\citenamefont{Stingelin-Stutzmann, Smits, Wondergem, Tanase,
  Blom, Smith, and DeLeeuw}}]{Stutzmann05}
\bibinfo{author}{\bibfnamefont{N.}~\bibnamefont{Stingelin-Stutzmann}},
  \bibinfo{author}{\bibfnamefont{E.}~\bibnamefont{Smits}},
  \bibinfo{author}{\bibfnamefont{H.}~\bibnamefont{Wondergem}},
  \bibinfo{author}{\bibfnamefont{C.}~\bibnamefont{Tanase}},
  \bibinfo{author}{\bibfnamefont{P.}~\bibnamefont{Blom}},
  \bibinfo{author}{\bibfnamefont{P.}~\bibnamefont{Smith}}, \bibnamefont{and}
  \bibinfo{author}{\bibfnamefont{D.}~\bibnamefont{DeLeeuw}},
  \bibinfo{journal}{Nature Mat.} \textbf{\bibinfo{volume}{4}},
  \bibinfo{pages}{601} (\bibinfo{year}{2005}).

\bibitem[{\citenamefont{Pope and Swenberg}(1999)}]{pope}
\bibinfo{author}{\bibfnamefont{M.}~\bibnamefont{Pope}} \bibnamefont{and}
  \bibinfo{author}{\bibfnamefont{C.~E.} \bibnamefont{Swenberg}},
  \emph{\bibinfo{title}{Electronic Processes in Organic Crystals and Polymers}}
  (\bibinfo{publisher}{Oxford University Press}, \bibinfo{year}{1999}),
  \bibinfo{edition}{2nd} ed.

\bibitem[{\citenamefont{Curtis et~al.}(2004)\citenamefont{Curtis, Cao, and
  Kampf}}]{Curtis2004}
\bibinfo{author}{\bibfnamefont{M.~C.} \bibnamefont{Curtis}},
  \bibinfo{author}{\bibfnamefont{J.}~\bibnamefont{Cao}}, \bibnamefont{and}
  \bibinfo{author}{\bibfnamefont{J.~W.} \bibnamefont{Kampf}},
  \bibinfo{journal}{J. Am. Chem. Soc.} \textbf{\bibinfo{volume}{126}},
  \bibinfo{pages}{4318} (\bibinfo{year}{2004}).

\bibitem[{\citenamefont{Anthony et~al.}(2001)\citenamefont{Anthony, Brooks,
  Eaton, and Parkin}}]{Anthony2001}
\bibinfo{author}{\bibfnamefont{J.~E.} \bibnamefont{Anthony}},
  \bibinfo{author}{\bibfnamefont{J.~S.} \bibnamefont{Brooks}},
  \bibinfo{author}{\bibfnamefont{D.~L.} \bibnamefont{Eaton}}, \bibnamefont{and}
  \bibinfo{author}{\bibfnamefont{S.~R.} \bibnamefont{Parkin}},
  \bibinfo{journal}{J. Am. Chem. Soc.} \textbf{\bibinfo{volume}{123}},
  \bibinfo{pages}{9482} (\bibinfo{year}{2001}).

\bibitem[{\citenamefont{Payne et~al.}(2005)\citenamefont{Payne, Parkin,
  Anthony, Kuo, and Jackson}}]{Payne2005a}
\bibinfo{author}{\bibfnamefont{M.~M.} \bibnamefont{Payne}},
  \bibinfo{author}{\bibfnamefont{S.~R.} \bibnamefont{Parkin}},
  \bibinfo{author}{\bibfnamefont{J.~E.} \bibnamefont{Anthony}},
  \bibinfo{author}{\bibfnamefont{C.-C.} \bibnamefont{Kuo}}, \bibnamefont{and}
  \bibinfo{author}{\bibfnamefont{T.~N.} \bibnamefont{Jackson}},
  \bibinfo{journal}{J. Am. Chem. Soc.} \textbf{\bibinfo{volume}{127}},
  \bibinfo{pages}{4986} (\bibinfo{year}{2005}).

\bibitem[{\citenamefont{Mas-Torrent et~al.}(2004)\citenamefont{Mas-Torrent,
  Hadley, Bromley, Ribas, Tarr\'es, Mas, Molins, Veciana, and
  Rovira}}]{Mas-Torrent2004}
\bibinfo{author}{\bibfnamefont{M.}~\bibnamefont{Mas-Torrent}},
  \bibinfo{author}{\bibfnamefont{P.}~\bibnamefont{Hadley}},
  \bibinfo{author}{\bibfnamefont{S.~T.} \bibnamefont{Bromley}},
  \bibinfo{author}{\bibfnamefont{X.}~\bibnamefont{Ribas}},
  \bibinfo{author}{\bibfnamefont{J.}~\bibnamefont{Tarr\'es}},
  \bibinfo{author}{\bibfnamefont{M.}~\bibnamefont{Mas}},
  \bibinfo{author}{\bibfnamefont{E.}~\bibnamefont{Molins}},
  \bibinfo{author}{\bibfnamefont{J.}~\bibnamefont{Veciana}}, \bibnamefont{and}
  \bibinfo{author}{\bibfnamefont{C.}~\bibnamefont{Rovira}},
  \bibinfo{journal}{J. Am. Chem. Soc.} \textbf{\bibinfo{volume}{126}},
  \bibinfo{pages}{8546} (\bibinfo{year}{2004}).

\bibitem[{\citenamefont{Moon et~al.}(2004)\citenamefont{Moon, Zeis, Borkent,
  Besnard, Lovinger, Siegrist, Kloc, and Bao}}]{TcD}
\bibinfo{author}{\bibfnamefont{H.}~\bibnamefont{Moon}},
  \bibinfo{author}{\bibfnamefont{R.}~\bibnamefont{Zeis}},
  \bibinfo{author}{\bibfnamefont{E.-J.} \bibnamefont{Borkent}},
  \bibinfo{author}{\bibfnamefont{C.}~\bibnamefont{Besnard}},
  \bibinfo{author}{\bibfnamefont{A.~J.} \bibnamefont{Lovinger}},
  \bibinfo{author}{\bibfnamefont{T.}~\bibnamefont{Siegrist}},
  \bibinfo{author}{\bibfnamefont{C.}~\bibnamefont{Kloc}}, \bibnamefont{and}
  \bibinfo{author}{\bibfnamefont{Z.}~\bibnamefont{Bao}}, \bibinfo{journal}{J.
  Am. Chem. Soc} \textbf{\bibinfo{volume}{126}}, \bibinfo{pages}{15322}
  (\bibinfo{year}{2004}).

\bibitem[{\citenamefont{Kopranenkov and Luk'yanets}(1972)}]{Kopranenkov1972}
\bibinfo{author}{\bibfnamefont{V.~N.} \bibnamefont{Kopranenkov}}
  \bibnamefont{and} \bibinfo{author}{\bibfnamefont{E.~A.}
  \bibnamefont{Luk'yanets}}, \bibinfo{journal}{Zh. Org. Khim.}
  \textbf{\bibinfo{volume}{8}}, \bibinfo{pages}{1690} (\bibinfo{year}{1972}).

\bibitem[{\citenamefont{Kloc et~al.}(1997)\citenamefont{Kloc, Simpkins,
  Siegrist, and Laudise}}]{Kloc1997}
\bibinfo{author}{\bibfnamefont{C.}~\bibnamefont{Kloc}},
  \bibinfo{author}{\bibfnamefont{P.~G.} \bibnamefont{Simpkins}},
  \bibinfo{author}{\bibfnamefont{T.}~\bibnamefont{Siegrist}}, \bibnamefont{and}
  \bibinfo{author}{\bibfnamefont{R.~A.} \bibnamefont{Laudise}},
  \bibinfo{journal}{J. Cryst. Growth} \textbf{\bibinfo{volume}{182(3-4)}},
  \bibinfo{pages}{416} (\bibinfo{year}{1997}).

\bibitem[{\citenamefont{Laudise et~al.}(1998)\citenamefont{Laudise, Kloc,
  Simpkins, and Siegrist}}]{Laudise1998}
\bibinfo{author}{\bibfnamefont{R.~A.} \bibnamefont{Laudise}},
  \bibinfo{author}{\bibfnamefont{C.}~\bibnamefont{Kloc}},
  \bibinfo{author}{\bibfnamefont{P.~G.} \bibnamefont{Simpkins}},
  \bibnamefont{and} \bibinfo{author}{\bibfnamefont{T.}~\bibnamefont{Siegrist}},
  \bibinfo{journal}{J. Cryst. Growth} \textbf{\bibinfo{volume}{187(3-4)}},
  \bibinfo{pages}{449} (\bibinfo{year}{1998}).

\bibitem[{\citenamefont{Farrugia}(1997)}]{Farrugia1997}
\bibinfo{author}{\bibfnamefont{L.~J.} \bibnamefont{Farrugia}},
  \bibinfo{journal}{J. Appl. Cryst.} \textbf{\bibinfo{volume}{30}},
  \bibinfo{pages}{565} (\bibinfo{year}{1997}).

\bibitem[{\citenamefont{Takeya et~al.}(2003)\citenamefont{Takeya, Goldmann,
  Haas, Pernstich, Ketterer, and Batlogg}}]{Takeya2003}
\bibinfo{author}{\bibfnamefont{J.}~\bibnamefont{Takeya}},
  \bibinfo{author}{\bibfnamefont{C.}~\bibnamefont{Goldmann}},
  \bibinfo{author}{\bibfnamefont{S.}~\bibnamefont{Haas}},
  \bibinfo{author}{\bibfnamefont{K.~P.} \bibnamefont{Pernstich}},
  \bibinfo{author}{\bibfnamefont{B.}~\bibnamefont{Ketterer}}, \bibnamefont{and}
  \bibinfo{author}{\bibfnamefont{B.}~\bibnamefont{Batlogg}},
  \bibinfo{journal}{J. Appl. Phys.} \textbf{\bibinfo{volume}{94(9)}},
  \bibinfo{pages}{5800} (\bibinfo{year}{2003}).

\bibitem[{\citenamefont{Gundlach et~al.}(2001)\citenamefont{Gundlach, Jia, and
  Jackson}}]{Gundlach2001a}
\bibinfo{author}{\bibfnamefont{D.~J.} \bibnamefont{Gundlach}},
  \bibinfo{author}{\bibfnamefont{L.~L.} \bibnamefont{Jia}}, \bibnamefont{and}
  \bibinfo{author}{\bibfnamefont{T.~N.} \bibnamefont{Jackson}},
  \bibinfo{journal}{IEEE Electron Device Lett.} \textbf{\bibinfo{volume}{22}},
  \bibinfo{pages}{571} (\bibinfo{year}{2001}).

\bibitem[{\citenamefont{Krellner et~al.}(2007)\citenamefont{Krellner, Haas,
  Goldmann, Pernstich, Gundlach, and Batlogg}}]{Krellner}
\bibinfo{author}{\bibfnamefont{C.}~\bibnamefont{Krellner}},
  \bibinfo{author}{\bibfnamefont{S.}~\bibnamefont{Haas}},
  \bibinfo{author}{\bibfnamefont{C.}~\bibnamefont{Goldmann}},
  \bibinfo{author}{\bibfnamefont{K.~P.} \bibnamefont{Pernstich}},
  \bibinfo{author}{\bibfnamefont{D.~J.} \bibnamefont{Gundlach}},
  \bibnamefont{and} \bibinfo{author}{\bibfnamefont{B.}~\bibnamefont{Batlogg}},
  \bibinfo{journal}{Phys. Rev. B} \textbf{\bibinfo{volume}{75}},
  \bibinfo{pages}{245115} (\bibinfo{year}{2007}).

\bibitem[{\citenamefont{Bulgarovskaya et~al.}(1983)\citenamefont{Bulgarovskaya,
  Vozzhennikov, Aleksandrov, and Belsky}}]{Bulgarovskaya83}
\bibinfo{author}{\bibfnamefont{I.}~\bibnamefont{Bulgarovskaya}},
  \bibinfo{author}{\bibfnamefont{V.}~\bibnamefont{Vozzhennikov}},
  \bibinfo{author}{\bibfnamefont{S.}~\bibnamefont{Aleksandrov}},
  \bibnamefont{and} \bibinfo{author}{\bibfnamefont{V.}~\bibnamefont{Belsky}},
  \bibinfo{journal}{Latv. PSR Zinat. Akad. Vestis, Khim. Ser.}
  \textbf{\bibinfo{volume}{4}}, \bibinfo{pages}{53} (\bibinfo{year}{1983}).

\bibitem[{\citenamefont{Schuck et~al.}(2007{\natexlab{a}})\citenamefont{Schuck,
  Haas, Stassen, and Batlogg}}]{schuck}
\bibinfo{author}{\bibfnamefont{G.}~\bibnamefont{Schuck}},
  \bibinfo{author}{\bibfnamefont{S.}~\bibnamefont{Haas}},
  \bibinfo{author}{\bibfnamefont{A.}~\bibnamefont{Stassen}}, \bibnamefont{and}
  \bibinfo{author}{\bibfnamefont{B.}~\bibnamefont{Batlogg}},
  \bibinfo{journal}{Acta Crystallogr. E} \textbf{\bibinfo{volume}{63}},
  \bibinfo{pages}{o2894} (\bibinfo{year}{2007}{\natexlab{a}}).

\bibitem[{\citenamefont{Mattheus et~al.}(2001)\citenamefont{Mattheus, Dros,
  Baas, Meetsma, de~Boer, and Palstra}}]{Mattheus2001}
\bibinfo{author}{\bibfnamefont{C.~C.} \bibnamefont{Mattheus}},
  \bibinfo{author}{\bibfnamefont{A.~B.} \bibnamefont{Dros}},
  \bibinfo{author}{\bibfnamefont{J.}~\bibnamefont{Baas}},
  \bibinfo{author}{\bibfnamefont{A.}~\bibnamefont{Meetsma}},
  \bibinfo{author}{\bibfnamefont{J.~L.} \bibnamefont{de~Boer}},
  \bibnamefont{and} \bibinfo{author}{\bibfnamefont{T.~T.~M.}
  \bibnamefont{Palstra}}, \bibinfo{journal}{Acta Crystallogr. C}
  \textbf{\bibinfo{volume}{57}}, \bibinfo{pages}{939} (\bibinfo{year}{2001}).

\bibitem[{RuD({\natexlab{a}})}]{RuD_RubreneAchsen}
\bibinfo{note}{Bulgarovskaya et al. \cite{Bulgarovskaya83} use a different
  space group setting ($Bbam$) than e.g. Sundar et al. \cite{Sundar2004}
  ($Acam$), resulting in the interchange of $a$ and $b$.}

\bibitem[{\citenamefont{Schuck et~al.}(2007{\natexlab{b}})\citenamefont{Schuck,
  Haas, Stassen, and Batlogg}}]{schuck1}
\bibinfo{author}{\bibfnamefont{G.}~\bibnamefont{Schuck}},
  \bibinfo{author}{\bibfnamefont{S.}~\bibnamefont{Haas}},
  \bibinfo{author}{\bibfnamefont{A.~F.} \bibnamefont{Stassen}},
  \bibnamefont{and} \bibinfo{author}{\bibfnamefont{B.}~\bibnamefont{Batlogg}},
  \bibinfo{journal}{Acta Crystallogr. E} \textbf{\bibinfo{volume}{63}},
  \bibinfo{pages}{o2893} (\bibinfo{year}{2007}{\natexlab{b}}).

\bibitem[{\citenamefont{Stassen et~al.}()\citenamefont{Stassen, Haas, Schuck,
  and Batlogg}}]{Arno_RuD}
\bibinfo{author}{\bibfnamefont{A.~F.} \bibnamefont{Stassen}},
  \bibinfo{author}{\bibfnamefont{S.}~\bibnamefont{Haas}},
  \bibinfo{author}{\bibfnamefont{G.}~\bibnamefont{Schuck}}, \bibnamefont{and}
  \bibinfo{author}{\bibfnamefont{B.}~\bibnamefont{Batlogg}}, \bibinfo{note}{to
  be published}.

\bibitem[{RuD({\natexlab{b}})}]{RuD_NMR}
\bibinfo{note}{5,11-BTBR and 5,12-BTBR can be clearly distinguished by the
  respective $^{13}$C-NMR signals.}

\bibitem[{\citenamefont{Tiago et~al.}(2003)\citenamefont{Tiago, Northrup, and
  Louie}}]{Tiago2003}
\bibinfo{author}{\bibfnamefont{M.~L.} \bibnamefont{Tiago}},
  \bibinfo{author}{\bibfnamefont{J.~E.} \bibnamefont{Northrup}},
  \bibnamefont{and} \bibinfo{author}{\bibfnamefont{S.~G.} \bibnamefont{Louie}},
  \bibinfo{journal}{Phys. Rev. B} \textbf{\bibinfo{volume}{67}},
  \bibinfo{pages}{115212} (\bibinfo{year}{2003}).

\bibitem[{\citenamefont{Hummer and Ambrosch-Draxl}(2005)}]{Hummer2005}
\bibinfo{author}{\bibfnamefont{K.}~\bibnamefont{Hummer}} \bibnamefont{and}
  \bibinfo{author}{\bibfnamefont{C.}~\bibnamefont{Ambrosch-Draxl}},
  \bibinfo{journal}{Phys. Rev. B} \textbf{\bibinfo{volume}{72}},
  \bibinfo{pages}{205205} (\bibinfo{year}{2005}).

\bibitem[{\citenamefont{Norton and Houk}(2005)}]{Norton2005}
\bibinfo{author}{\bibfnamefont{J.~E.} \bibnamefont{Norton}} \bibnamefont{and}
  \bibinfo{author}{\bibfnamefont{K.~N.} \bibnamefont{Houk}},
  \bibinfo{journal}{J. Am. Chem. Soc.} \textbf{\bibinfo{volume}{127}},
  \bibinfo{pages}{4162} (\bibinfo{year}{2005}).

\bibitem[{\citenamefont{Lu et~al.}(2004)\citenamefont{Lu, Ho, Vogelaar, Kraml,
  and Pascal~Jr.}}]{Lu2004}
\bibinfo{author}{\bibfnamefont{J.}~\bibnamefont{Lu}},
  \bibinfo{author}{\bibfnamefont{D.~M.} \bibnamefont{Ho}},
  \bibinfo{author}{\bibfnamefont{N.~J.} \bibnamefont{Vogelaar}},
  \bibinfo{author}{\bibfnamefont{C.~M.} \bibnamefont{Kraml}}, \bibnamefont{and}
  \bibinfo{author}{\bibfnamefont{R.~A.} \bibnamefont{Pascal~Jr.}},
  \bibinfo{journal}{J. Am. Chem. Soc.} \textbf{\bibinfo{volume}{126}},
  \bibinfo{pages}{11168} (\bibinfo{year}{2004}).

\bibitem[{\citenamefont{Schauer et~al.}(1996)\citenamefont{Schauer,
  Ne\^sp$\mathring{\textnormal{u}}$rek, and Valeri\'an}}]{Schauer04}
\bibinfo{author}{\bibfnamefont{F.}~\bibnamefont{Schauer}},
  \bibinfo{author}{\bibfnamefont{S.}~\bibnamefont{Ne\^sp$\mathring{\textnormal%
{u}}$rek}}, \bibnamefont{and}
  \bibinfo{author}{\bibfnamefont{H.}~\bibnamefont{Valeri\'an}},
  \bibinfo{journal}{J. Appl. Phys.} \textbf{\bibinfo{volume}{80}},
  \bibinfo{pages}{880} (\bibinfo{year}{1996}).

\bibitem[{\citenamefont{Lang et~al.}(2004)\citenamefont{Lang, Chi, Siegrist,
  Sergent, and Ramirez}}]{Lang02}
\bibinfo{author}{\bibfnamefont{D.~V.} \bibnamefont{Lang}},
  \bibinfo{author}{\bibfnamefont{X.}~\bibnamefont{Chi}},
  \bibinfo{author}{\bibfnamefont{T.}~\bibnamefont{Siegrist}},
  \bibinfo{author}{\bibfnamefont{A.}~\bibnamefont{Sergent}}, \bibnamefont{and}
  \bibinfo{author}{\bibfnamefont{A.~P.} \bibnamefont{Ramirez}},
  \bibinfo{journal}{Phys. Rev. Lett.} \textbf{\bibinfo{volume}{93(8)}},
  \bibinfo{pages}{086802} (\bibinfo{year}{2004}).

\bibitem[{\citenamefont{Krellner}(2004)}]{diplom}
\bibinfo{author}{\bibfnamefont{C.}~\bibnamefont{Krellner}},
  \emph{\bibinfo{title}{Diploma thesis}}, \bibinfo{howpublished}{{E}TH Zurich /
  University of Dresden} (\bibinfo{year}{2004}).

\end{thebibliography}
\end{document}